# CC-OTDR Sequence Shaping Enabling Joint Co-directional Sensing and Communication

M. Ali Allousch and André Sandmann

Advanced Technology, Adtran Networks SE, Maerzenquelle 1-3, 98617 Meiningen, Germany

Ali.Allousch@adtran.com

*Abstract*— CC-OTDR signal envelope shaping is introduced to reduce the impact of non-linear signal interactions on a neighboring wavelength data channel when co-propagating the probing signal with the data signal. Joint co-directional acoustic sensing and 200 Gbps transmission are demonstrated over a 50 km link.

*Keywords—CC-OTDR, Joint Communication and Sensing, sequence-shaping*

## I. INTRODUCTION

Joint communication and sensing (JCAS) is a key enabler for future intelligent networks, offering efficient exploitation of infrastructure by combining high-speed data transmission with real-time environmental monitoring [1-2]. Optical fibers serve as an ideal medium for JCAS due to their low loss, high bandwidth, and compatibility with both communication and sensing. However, the simultaneous propagation of sensing probes and coherent data signals in the same fiber introduces nonlinear signal interactions that can degrade the performance of both systems, posing a significant challenge for practical deployment. This novel work addresses the use of power envelope shaping of coherent correlation optical time-domain reflectometry (CC-OTDR) probe sequences to minimize the transmission performance penalty when co-propagating the probing signal with the data signal. Unlike conventional rectangular probe pulses or power envelopes, the shaped sequence avoids abrupt power transitions, mitigating the effect of phase jumps in the data signal from cross-phase modulation. This approach introduces a compatibility mode to enable in-band sensing in deployments where co-propagation of the probing signal is required with minimal penalty.

## II. EXPERIMENTAL SETUP

The experimental setup is illustrated in Fig. 1. JCAS is implemented in a co-directional configuration by launching a 193.4 THz optical probe from a CC-OTDR-based distributed acoustic sensing testbed [3-4], and a 193.5 THz dual-polarization coherent data signal from an optical transmitter (Tx) into a shared fiber. A symbol rate of 33 GBaud and QPSK as well as 16-QAM formats were tested. The transmitter output is amplified by a booster erbium-doped fiber amplifier (EDFA1) and attenuated by a variable optical attenuator (VOA1) to obtain a launch power of $P_{\text{avg}} = 0$ dBm after the wavelength multiplexer (MUX), which combines both signals. After the MUX, the CC-OTDR signal peak power is $P_{\text{peak}} = 7.8$ dBm. The combined signals propagate through a fiber link that also serves as the sensing medium. This link consists of an initial 8-meter fiber segment, followed by a 50 km fiber spool and a piezoelectric transducer (PZT) introducing a dynamic sinusoidal strain variation of 115 Hz. Both the spool and the PZT are placed inside an acoustically isolated chamber to suppress environmental noise and ensure controlled conditions. A final 8-meter fiber segment completes the sensing path. At the fiber end, a wavelength demultiplexer (DEMUX) separates the sensing and communication channels. This separation is essential to terminate the sensing path and avoid unintended amplification of the sensing signal, which could otherwise perturb the gain dynamics and degrade the quality of the communication signal. Subsequently, configurable amplified spontaneous emission (ASE) noise is added via a 3 dB coupler, and the noise level can be controlled via VOA2, emulating optical signal-to-noise ratio (OSNR) degradation of a multi-span transmission. The pre-amplifier EDFA2 and VOA3 are used to adjust the signal power in the optimal range for the receiver. Additionally, 10% of the signal is tapped and directed to an optical spectrum analyzer (OSA) for OSNR measurement, while the remaining 90% is delivered via an optical bandpass filter (OBF) to the receiver (Rx) for data recovery.

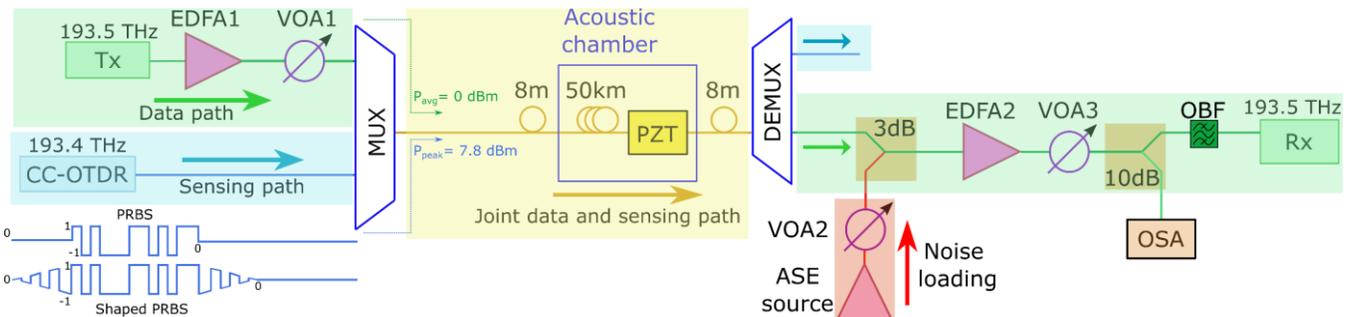

Fig. 1. Schematic of the experimental setup for co-directional joint communication and sensing using CC-OTDR and a 200 Gbps coherent transceiver. PRBS and shaped PRBS sequence illustrations are depicted (bottom-left).

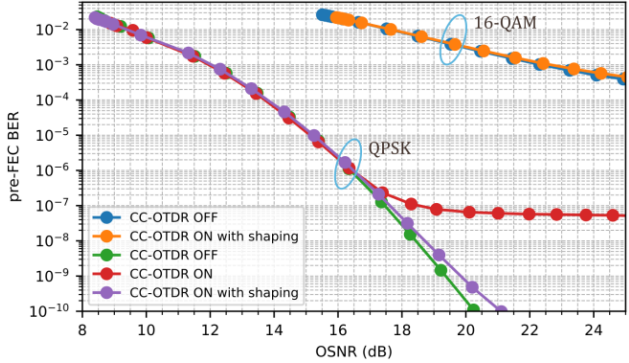

Fig. 2. Pre-FEC BER vs. OSNR for QPSK and 16-QAM under different CC-OTDR probing signal configurations.

TABLE I. REQUIRED OSNR FOR CO-DIRECTIONAL OPERATION AND 100 GHZ SPACING

| Mode | Required OSNR |
|---|---|
| 16-QAM CC-OTDR OFF | 15.5 dB |
| 16-QAM CC-OTDR ON | uncorrected blocks |
| 16-QAM CC-OTDR ON with shaping | 15.9 dB |
| QPSK CC-OTDR OFF | 8.4 dB |
| QPSK CC-OTDR ON | 8.8 dB |
| QPSK CC-OTDR ON with shaping | 8.4 dB |

## III. EXPERIMENTAL RESULTS

The impact of the CC-OTDR system on the data transmission quality was evaluated using two code sequences: a 4095-bit pseudo-random binary sequence (PRBS) balanced by adding a "-1" symbol and a corresponding trapezoidal-shaped 8192-bit sequence, which embeds the same PRBS centrally with 2048-bit ramp-up and ramp-down segments on either side. The impact of both sequences on the transmission performance was tested for QPSK and 16-QAM. Fig. 2 shows the pre-forward error correction (FEC) bit-error rate (BER) versus OSNR for QPSK and 16-QAM under three conditions: CC-OTDR off, CC-OTDR on with conventional PRBS, and CC-OTDR on with the shaped sequence. For QPSK, a BER floor at $5.2 \cdot 10^{-8}$ is visible at high OSNR levels when CC-OTDR is switched on. With shaping, the BER approaches the level observed when CC-OTDR is switched off. The required OSNR to obtain quasi error-free transmission after FEC decoding is shown in Table I. It shows a penalty of 0.4 dB when no CC-OTDR signal shaping is used. With signal shaping, the required OSNR is similar to the case with CC-OTDR off. In contrast, 16-QAM transmission without post-FEC errors is only possible with the use of signal shaping. Without signal shaping, uncorrected blocks frequently occurred even when noise loading was disabled. With signal shaping, the BER performance closely matches the case when CC-OTDR is switched off, and the corresponding required OSNR penalty is just 0.4 dB. The abrupt power transitions at the start and end of the PRBS sequence or any pulse-based sensing signal cause phase jumps in the co-propagating data signal from cross-phase modulation. It is assumed that the commercial digital signal processor used cannot compensate for these phase jumps, which results in burst errors. In contrast, the smoother power transitions of a shaped sequence avoid this issue, enabling interoperability of communication and sensing in scenarios where co-propagation of signals is required.

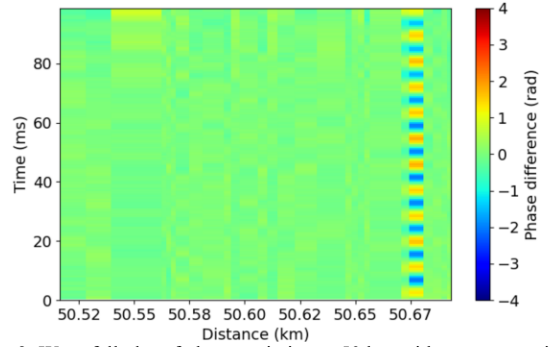

Fig. 3. Waterfall plot of phase variation at 50 km with co-propagating 16-QAM data signals with shaped CC-OTDR sequence.

To verify sensing performance with signal shaping while transmitting data with 16-QAM, a 115 Hz, 125 mV peak-to-peak signal was applied to the piezoelectric transducer (PZT). The resulting phase variation with an amplitude of 3.9 rad peak-to-peak at 50.67 km is shown in the waterfall plot in Fig. 3, confirming the CC-OTDR's capability for distributed acoustic sensing alongside a co-propagating 200 Gbps data channel.

## IV. CONCLUSION

This work demonstrates a novel approach to enabling in-band sensing by introducing signal shaping of CC-OTDR probe sequences for smooth interoperability with co-propagating 200 Gbps data signals at a frequency spacing of 100 GHz. The shaped waveform effectively mitigates phase jumps in the data signal from cross-phase modulation, overcoming limitations observed with PRBS sequence or other pulse-based sensing probe signals. Experimental results confirm that this shaping strategy allows reliable coherent data transmission with minimal BER performance penalty alongside distributed acoustic sensing on a shared optical fiber, marking a significant step toward practical and scalable JCAS.

## ACKNOWLEDGMENT

This work has received funding from the Horizon Europe Framework Programme under grant agreement No 101189654 (GASPOF Project).